\documentclass[sigplan,10pt]{acmart}
\renewcommand\footnotetextcopyrightpermission[1]{}
\usepackage{tikz}
\usepackage{amsmath}
\usepackage{xspace}
\usepackage{enumitem}
\usepackage{subcaption}
\usepackage{nicefrac}
\usepackage{soul}

\usepackage[ruled, linesnumbered, vlined]{algorithm2e}
\DontPrintSemicolon

\SetCommentSty{mycommfont}

\PassOptionsToPackage{hyphens}{url}

\newcommand{\sys}{\textsc{PASTE}\xspace}
\newcommand{\authorformat}[1]{\textbf{\Large #1}}
\newcommand{\affilformat}[1]{\textit{\large #1}}

\makeatletter
\def\@mkauthors{%
  \global\setbox\mktitle@bx=\vbox{%
    \noindent\unvbox\mktitle@bx
    \par\medskip
    \begin{center}
      \authorformat{Yifan Sui\textsuperscript{1}}\quad
      \authorformat{Han Zhao\textsuperscript{1}}\quad
      \authorformat{Rui Ma\textsuperscript{2}}\quad
      \authorformat{Zhiyuan He\textsuperscript{2}}\quad
      \authorformat{Hao Wang\textsuperscript{3}}\par
      \authorformat{Kaiqiang Xu\textsuperscript{4,5}}\quad
      \authorformat{Kai Chen\textsuperscript{5}}\quad
      \authorformat{Jianxun Li\textsuperscript{1}}\quad
      \authorformat{Yuqing Yang\textsuperscript{2}}\par
      \vspace{1.2em}
      \affilformat{\textsuperscript{1}Shanghai Jiao Tong University}\quad
      \affilformat{\textsuperscript{2}Microsoft Research}\quad
      \affilformat{\textsuperscript{3}Stevens Institute of Technology}\par
      \affilformat{\textsuperscript{4}Google}\quad
      \affilformat{\textsuperscript{5}HKUST}
    \end{center}
    \par\bigskip
  }%
}
\makeatother

\acmConference[arXiv preprint]{}{2026}{}\acmYear{2026}
\begin{document}

\title{
Parallelizing Tool Execution and LLM Generation for Low-Latency Agent Serving
}

\author{Yifan Sui}
\email{suiyifan@sjtu.edu.cn}
\affiliation{
  \institution{Shanghai Jiao Tong University}
  \city{Shanghai}
  \country{China}
}

\author{Han Zhao}
\email{zhaohan_miven@sjtu.edu.cn}
\affiliation{
  \institution{Shanghai Jiao Tong University}
  \city{Shanghai}
  \country{China}
}

\author{Rui Ma}
\email{mrui@microsoft.com}
\affiliation{
  \institution{Microsoft Research}
  \city{Beijing}
  \country{China}
}

\author{Zhiyuan He}
\email{zhiyuhe@microsoft.com}
\affiliation{
  \institution{Microsoft Research}
  \city{Beijing}
  \country{China}
}

\author{Hao Wang}
\email{hwang9@stevens.edu}
\affiliation{
  \institution{Stevens Institute of Technology}
  \city{Hoboken}
  \country{USA}
}

\author{Kaiqiang Xu}
\email{xkq@cse.ust.hk}
\affiliation{
  \institution{Google}
  \city{Mountain View}
  \country{USA}
}
\affiliation{
  \institution{HKUST}
  \city{Hong Kong}
  \country{China}
}

\author{Kai Chen}
\email{kaichen@cse.ust.hk}
\affiliation{
  \institution{Hong Kong University of Science and Technology}
  \city{Hong Kong}
  \country{China}
}

\author{Jianxun Li}
\email{lijx@sjtu.edu.cn}
\affiliation{
  \institution{Shanghai Jiao Tong University}
  \city{Shanghai}
  \country{China}
}

\author{Yuqing Yang}
\email{yuqyang@microsoft.com}
\affiliation{
  \institution{Microsoft Research}
  \city{Redmond}
  \country{USA}
}

\begin{abstract}



LLM-powered agents execute tasks through a sequential loop of model generation and tool execution. Today's serving systems serialize this loop, leaving tool latency exposed on the task critical path. 
This paper presents PASTE, a tool-aware agent-serving system that predicts concrete future tool invocations from recurring agent patterns and executes them speculatively while the LLM is still generating. PASTE isolates speculative results until confirmed by the LLM and jointly schedules tool execution and returning LLM sessions to avoid shifting bottlenecks to the GPU. Across deep research, coding, and scientific-agent workloads, PASTE reduces average task completion time by 43.5\% and lowers observed tool latency by 1.8$\times$. 
\end{abstract}

\maketitle
\def\v2#1{\textcolor{black}{#1}}

\section{Introduction}

\begin{figure}[t]
    \centering
    \includegraphics[width=0.98\columnwidth]{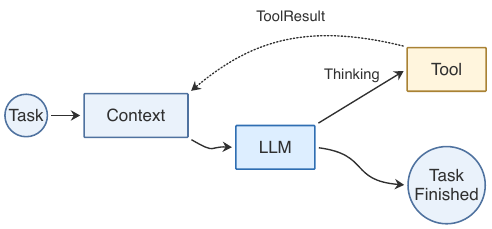}
    \vspace{-0.2in}
    \caption{Workflow of an LLM agent.}
    \label{fig:workflow}
    \vspace{-0.2in}
\end{figure}

LLM-powered agents are becoming a common execution model for long-running tasks such as deep research, software debugging, and scientific analysis~\cite{adminKimiResearcherKimi2025, GeminiCLIBuild2025, ClaudeCodeClaude, GitHubCopilotYour2025, claudeskills}. As exemplified by OpenAI Deep Research~\cite{IntroducingDeepResearch2025} and Manus~\cite{ManusHandsAI}, the LLM acts as the ``brain'' of the agent: it decomposes a complex task into sub-steps and invokes external tools to interact with the outside world~\cite{luo2025autellix, zhang2026megaflow, giusti2025federation}.

A modern LLM-powered agent follows an \emph{Iterative LLM--Tool Loop}, as shown in Fig.~\ref{fig:workflow}. \v2{Agent execution is semantically sequential: each step depends on the output of the previous step. In today's agent systems like Codex and Qwen Code, however, this semantic order is typically realized as a physically serialized path: the LLM generates, emits a tool call, waits for the tool result, and then resumes generation.} As a result, LLM generation and tool execution form a single critical path. Therefore, the agent serving system's latency objective shifts from minimizing token-level latency to minimizing task-level end-to-end (E2E) latency.

This execution model makes task-level E2E latency the central objective for agent serving, and exposed tool time a task-level bottleneck. Two observations motivate our work:

First, tool execution is a major part of the critical path: in our measurements (\S\ref{sec:inefficiencies}; Fig.~\ref{fig:moti:single}), it accounts for 45\%--57\% of agent E2E latency across representative deep research, coding, and scientific tasks. Because this tool portion lies on the serialized critical path, reducing task E2E latency requires reducing the exposed tool time, not just accelerating LLM generation. To move tool execution earlier without changing semantic order, the system must know concrete future tool invocations before they appear on the serialized path. 

\begin{figure}[t]
    \centering
    \includegraphics[width=\linewidth]{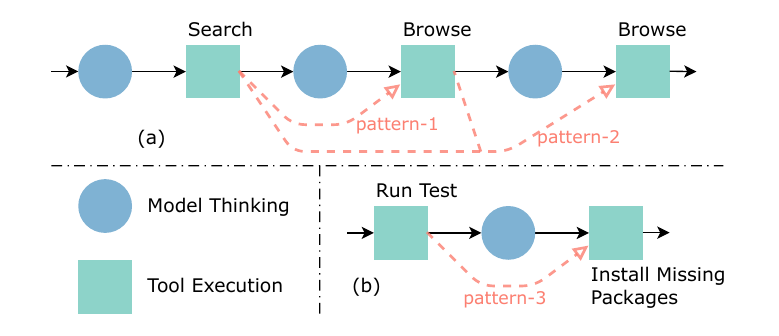}
    \vspace{-0.15in}
    \caption{Example recurring tool-call patterns and latent data dependencies in agent traces.}
    \label{fig:motifs}
    \vspace{-0.15in}
\end{figure}

Second, future tool use is often predictable before the LLM explicitly emits the next tool call. As Fig.~\ref{fig:motifs} shows, real traces show recurring sub-workflow patterns, such as \textit{search} followed by \textit{visit}, \textit{edit} followed by \textit{run test}, \textit{grep} followed by \textit{file\_editor}. These patterns often carry implicit data dependencies, where subsequent tool call's inputs are already contained in earlier calls' outputs. For example, 55\% of successful \textit{file\_editor} calls are followed by terminal execution, and in 95\% of web visits the URL is a strict substring of the preceding search output (\S\ref{subsec:opportunities}). \v2{Therefore, future tool invocations can often be predicted before they appear on the serialized path.}

\v2{Together, these two observations create an opportunity for serving-level parallelization: the system can reduce substantial exposed tool time by predicting concrete future tool invocations and executing them while the LLM is still generating. Instead of waiting for the LLM to emit every tool call, we can execute these invocations speculatively. This speculative execution creates temporal overlap between the tool execution and LLM generation. Through this overlap, the serving system parallelizes tool execution and LLM generation without changing the agent's semantically sequential workflow.}

However, to fully take advantage of this opportunity, we must address two challenges. First, effective speculation must predict concrete future invocations under uncertainty, not only tool names. The same pattern can imply different arguments in different sessions, and wrong arguments waste speculative work because the speculative results are not reusable when the authoritative invocation arrives. \v2{Second, blind speculation can create LLM--tool phase imbalance. Uncoordinated parallelization does not always reduce task-level latency:} correct speculation \v2{shortens exposed tool time, but it can also make many sessions become ready for LLM generation earlier and closer together, shifting the bottleneck from the tool phase to the LLM phase.} \v2{Because LLM generation is load-sensitive (\S\ref{subsec:observation2}), this phase imbalance could increase decoding time or queueing and erase the tool-side gain. On the other hand, inaccurate speculation further creates speculative results the LLM never consumes.} The system must jointly decide when to execute speculative tool calls and how to pace sessions that return from tool-side into the LLM batch, avoiding phase imbalance that converts tool acceleration into LLM slowdown.

Existing LLM-serving and agentic-serving systems are tool-unaware and therefore do not address the above challenges. General LLM serving systems are organized around token-level serving efficiency and schedule ready LLM requests as the primary workload~\cite{kwon2023efficient, agrawal2024sarathiserve, duan2024muxserve, qiao2024conserve}. Agentic-serving systems add session-aware mechanisms, but still treat a tool call as an external waiting interval between two LLM requests~\cite{abhyankar2024infercept, wang2025augserve, pan2025kvflow, xie2025strata, li2025throughput}. As a result, \v2{these systems cannot jointly optimize tool execution and LLM generation.} Because tool execution sits outside their scheduling loop, they leave tool time exposed on the task critical path; because they lack visibility into tool returns, they cannot pace sessions before they re-enter the LLM engine. Even if another mechanism shortens tool latency, the saved time can reappear as bursty LLM arrivals that push the GPU away from its latency-efficient region.


To address these challenges, we propose \textbf{P}attern-\textbf{A}ware \textbf{S}peculative \textbf{T}ool \textbf{E}xecution (\textbf{\sys}), a coordinated tool--LLM parallelization system for agent serving. \sys exploits recurring agent patterns to move concrete tool execution earlier in physical time, while preserving the agent's semantically sequential execution. Instead of treating the LLM engine and tool executor as independent components, \sys places a control plane across both phases: it speculates on likely future tool calls, isolates speculative results until they are confirmed by the LLM, and controls when completed sessions re-enter the LLM engine.


\sys consists of three core components: the Pattern Analyzer, the Tool Speculation Scheduler, and the LLM--Tool Co-Scheduler. The Pattern Analyzer mines recurring control flow and implicit data flow from prior agent traces, then instantiates concrete future tool invocations from live session state. The Tool Speculation Scheduler coordinates authoritative and speculative invocations. It admits speculative work when it is likely to hide exposed tool time, isolates speculative results until confirmation, and preserves priority for authoritative tool calls. The LLM--Tool Co-Scheduler paces returning sessions to keep the LLM engine away from both under-utilization and overload. Together, these components turn predictable tool waits into effective overlap with LLM generation, yielding task-level end-to-end speedup without changing agent semantics.

We implement \sys as a full-stack system spanning agent applications and the vLLM serving stack, covering both tool-side execution and LLM-side scheduling. Our evaluation demonstrates that compared to state-of-the-art baselines, \sys reduces the average task completion time by 43.5\% and speeds up average tool execution by 1.8$\times$.

The main contributions of this paper are as follows:

\begin{itemize}[itemsep=0pt, parsep=2pt, labelsep=5pt, leftmargin=*, topsep=2pt,partopsep=2pt]
\item \textbf{Characterization of the LLM--tool critical path.} We show that tool execution is a major critical-path component, and future tool invocations can often be predicted before they appear on the serialized path.
\item \textbf{Coordinated tool--LLM parallelization.} We propose \sys, which combines the Pattern Analyzer, the Tool Speculation Scheduler, and the LLM--Tool Co-Scheduler to turn recurring agent patterns into concrete predictions, execute speculative tool calls without interfering with authoritative execution, and preserve tool-side gains through LLM-side pacing.
\item \textbf{Full-stack implementation and evaluation.} We implement \sys across agent applications and the vLLM serving stack; compared to state-of-the-art baselines, \sys reduces the average task completion time by 43.5\% and speeds up average tool execution by 1.8$\times$.
\end{itemize}

\section{Background and Motivation}
\label{sec:moti}

\subsection{Agent Serving}

A modern LLM-powered agent follows an \emph{Iterative LLM--Tool Loop}, as shown in Fig.~\ref{fig:workflow}.
In each iteration, the LLM is conditioned on the current context and execution history, and then either emits the final answer or invokes a tool with concrete parameters.
If a tool is invoked, the tool executor runs the invocation, appends the observation back into the session state, and resumes LLM generation on the updated context.

This execution model changes the serving objective.
Unlike isolated LLM requests, agent sessions are long-running, stateful, and highly sequential.
Agent execution is semantically sequential: each step depends on the output of the previous step. In today's agent systems like Codex and Qwen Code, however, this semantic order is typically realized as a physically serialized path: the LLM generates, emits a tool call, waits for the tool result, and then resumes generation.
As a result, LLM generation and tool execution form a single critical path.
Therefore, the agent serving system's latency objective shifts from minimizing token-level latency to minimizing task-level end-to-end (E2E) latency across online sessions, including both average and tail latency.

\subsection{Observation 1: Tool Execution Is Critical-Path}
\label{sec:inefficiencies}

\begin{figure}[t]
    \centering
        \centering
        \includegraphics[width=\linewidth]{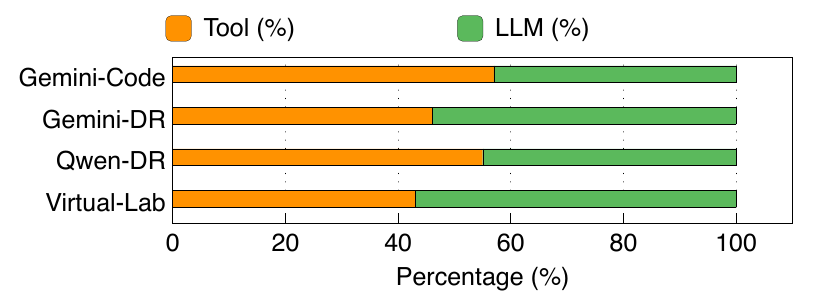}
    \vspace{-0.2in}
    \caption{Breakdown of tool execution and LLM generation.}
    \label{fig:moti:single}
    \vspace{-0.15in}
\end{figure}

We first measure the latency breakdown of individual agent requests under the most favorable serving condition: one agent request at a time, with no request queueing or resource contention.
We use representative agent workloads from deep research, coding, and AI-for-Science, and run Qwen-DeepResearch-30B on a server with 8 NVIDIA A100-80G GPUs.
As Fig.~\ref{fig:moti:single} shows, LLM generation and tool execution contribute comparable latency in this ideal setting.
Across the evaluated agents, tool execution accounts for 45\%--57\% of agent E2E latency.

This result shows that tool execution is substantial.
Because this semantic order is typically realized as a physically serialized path, the tool portion directly appears on the agent critical path.
Thus, reducing task E2E latency requires reducing the exposed tool time on this serialized path, not only accelerating LLM generation.

\subsection{Observation 2: Future Tool Use Is Predictable}
\label{subsec:opportunities}

\begin{figure}[t]
    \centering
    \includegraphics[width=0.98\linewidth]{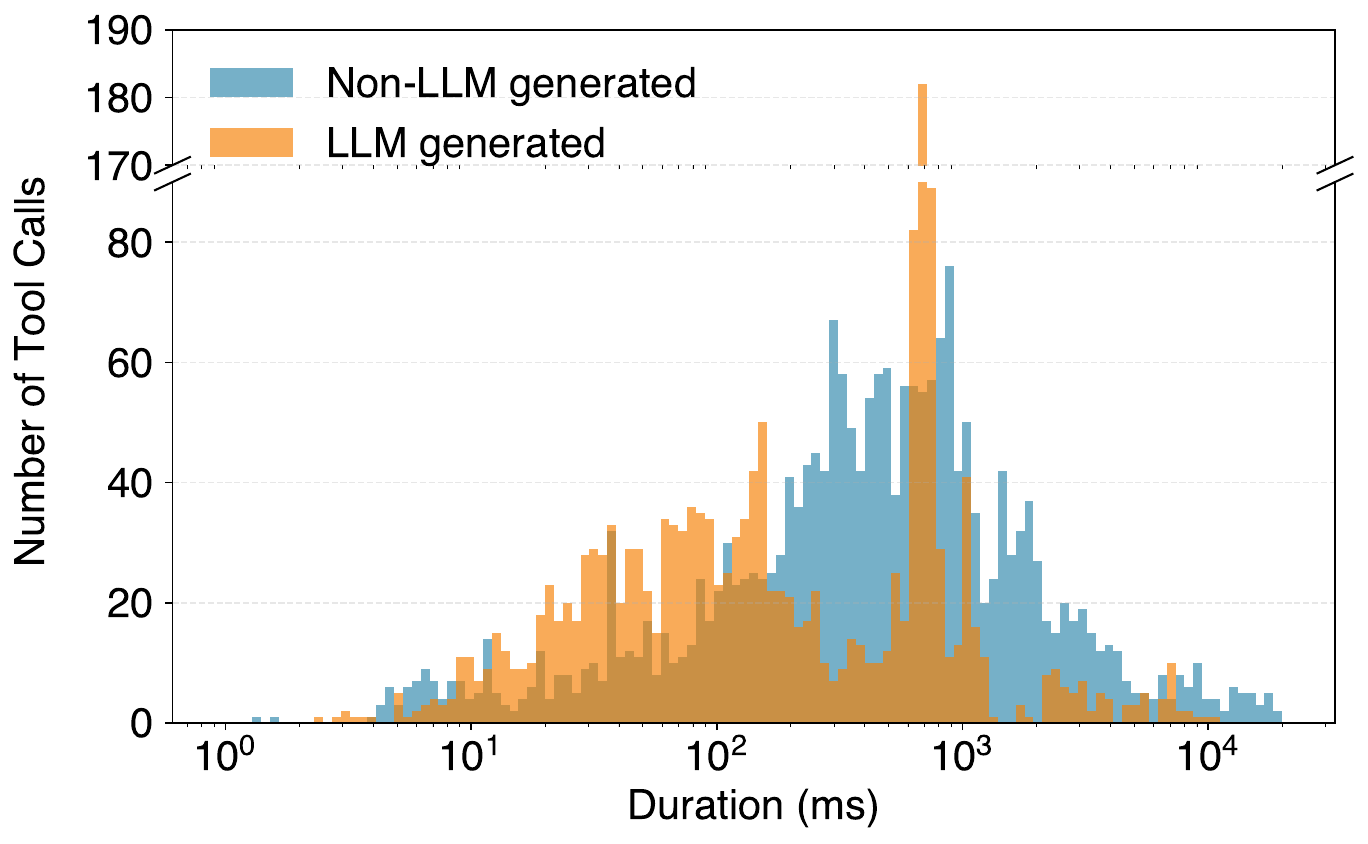}
    \vspace{-0.1in}
    \caption{Tool execution time histogram. Blue: args from prior outputs; orange: LLM-generated args.}
    \label{fig:moti:arg-source}
    \vspace{-0.1in}
\end{figure}

Reducing this exposed tool time without changing the agent's semantic order requires knowing concrete future invocations before they appear on the serialized path.
Future tool use is often predictable before the LLM explicitly emits the next tool call.
Agent tool use is generated online by the LLM.
It is not deterministic, but real traces still exhibit structure.
As Fig.~\ref{fig:motifs} shows, real traces show recurring sub-workflow patterns, such as \textit{search} followed by \textit{visit}, \textit{edit} followed by \textit{run test}, \textit{grep} followed by \textit{file\_editor}.
These patterns often carry implicit data dependencies, where subsequent tool call's inputs are already contained in earlier calls' outputs.
For example, 55\% of successful \textit{file\_editor} calls are followed by terminal execution, and in 95\% of web visits the URL is a strict substring of the preceding search output.
This predictability is not limited to tool names.
As Fig.~\ref{fig:moti:arg-source} shows, the measured tool calls are dominated by invocations whose arguments are derived from the prompt or outputs of previous tools rather than generated from scratch by the LLM in the current step.
These derived-argument tool calls are also latency-heavy, making them especially useful targets for reducing exposed tool time.
Therefore, future tool invocations can often be predicted before they appear on the serialized path.

\subsection{Opportunities and Challenges}
\label{subsec:observation2}

\textbf{Opportunity.} Together, these two observations create an opportunity for serving-level parallelization: the system can reduce substantial exposed tool time by predicting concrete future tool invocations and executing them while the LLM is still generating. Instead of waiting for the LLM to emit every tool call, \sys can execute these invocations speculatively. This speculative execution creates temporal overlap between the tool execution and LLM generation. Through this overlap, the serving system parallelizes tool execution and LLM generation without changing the agent's semantically sequential workflow.
The key is to move concrete tool execution earlier in physical time, not to change the agent's semantic order.
Once a predicted invocation can be instantiated from the observed context, the tool side can start executing it during the LLM generation interval that would otherwise precede the \textbf{authoritative tool call} (i.e., tool calls invoked by LLMs).
If the LLM later emits the same invocation, a completed speculative result can be reused, or an in-flight speculative execution can be promoted, so the session waits only for the remaining tool time instead of the full tool latency.
In this way, predictable control flow and implicit data flow become practical overlap between tool execution and LLM generation.

However, leveraging this opportunity presents two challenges:

\textbf{Challenge 1: Accurate and safe speculative tool execution.} Effective speculation must predict concrete future invocations under uncertainty, not only tool names.
The same pattern can imply different arguments in different sessions, and wrong arguments waste speculative work because the speculative results are not reusable when the authoritative invocation arrives.
A search may be followed by a visit, but the latency benefit only appears if the system chooses the right URL; an edit may be followed by a test, but the command and working directory must match the actual task.
These arguments are often copied or lightly transformed from earlier observations, so the system must recover implicit data-flow from noisy online context.
At the same time, speculative execution must be safe: predicted invocations should not delay, reorder, or otherwise affect authoritative invocations, and speculative actions that violate safety or side-effect constraints should not be admitted for execution.
Thus, the system needs both accurate concrete prediction and policy-compliant speculative execution before predicted tool invocations can be executed safely ahead of authoritative confirmation.

\begin{figure}[t]
    \centering
    \begin{subfigure}[b]{0.8\linewidth}
        \centering
        \includegraphics[width=\linewidth]{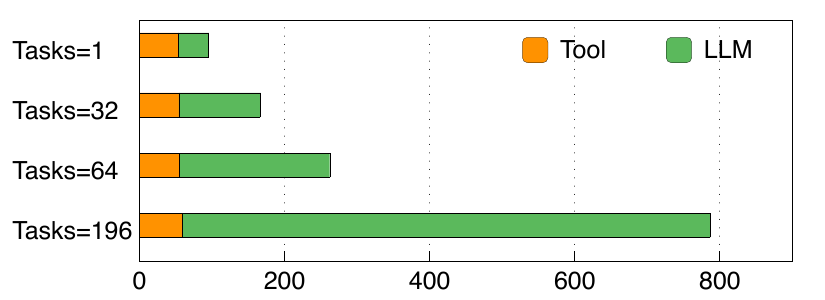}
        \caption{Absolute E2E time (s)}
        \label{fig:moti:conc-e2e}
    \end{subfigure}
    \begin{subfigure}[b]{0.8\linewidth}
        \centering
        \includegraphics[width=\linewidth]{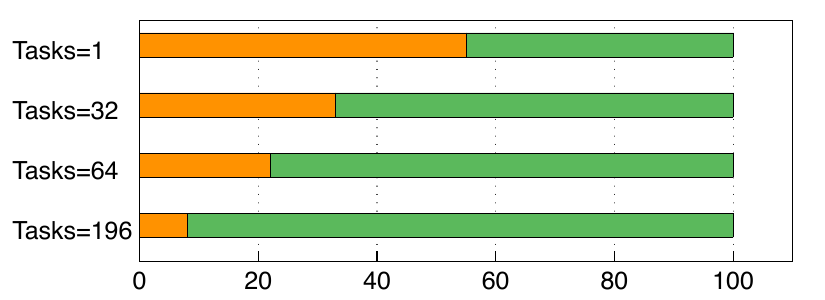}
        \caption{Time proportion (\%)}
        \label{fig:moti:conc-percent}
    \end{subfigure}
    \vspace{-0.1in}
    \caption{LLM generation and tool execution have different load sensitivity under concurrent agent tasks.}
    \label{fig:moti:concurrency}
    \vspace{-0.2in}
\end{figure}

\textbf{Challenge 2: Blind speculation can create LLM--tool phase imbalance.} The second challenge comes from the coupling between the tool phase and the LLM phase: LLM generation runs on shared GPUs and is highly sensitive to workload because its efficiency depends on dynamic resource utilization and memory constraints.
As Fig.~\ref{fig:moti:concurrency} shows, when concurrency grows from one to 192 concurrent sessions, LLM generation time grows by more than 17$\times$, while tool execution time changes only modestly.
Thus, any blind speculation policy that executes predicted tools as soon as possible and admits returning sessions immediately can make many sessions return to the LLM engine earlier and closer together, increasing active LLM load and shifting the bottleneck from the tool phase to the LLM phase.

\begin{figure}[t]
    \centering
    \includegraphics[width=\linewidth]{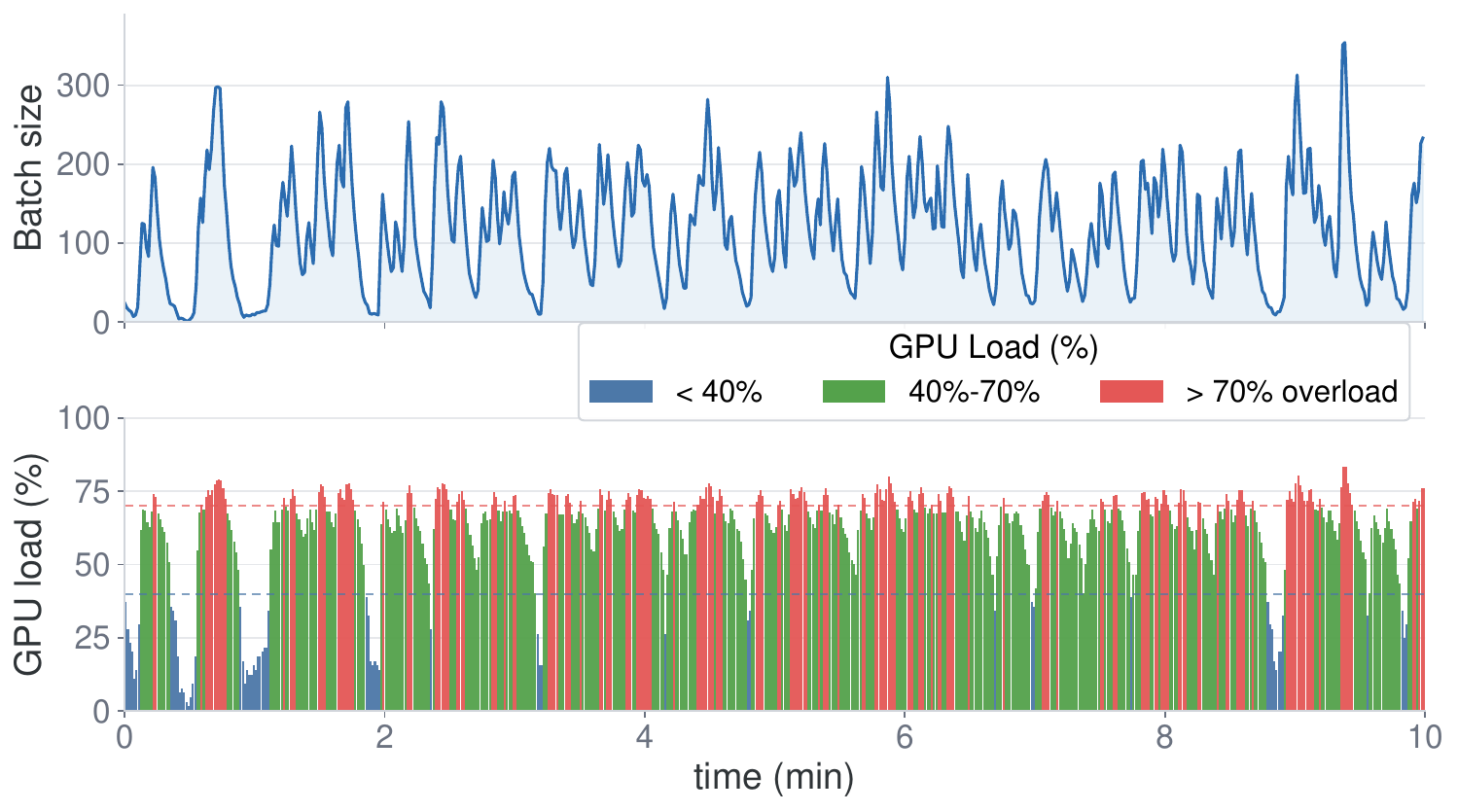}
    \vspace{-0.2in}
    \caption{Bursty GPU pressure from alternating LLM generation and tool execution.}
    \label{fig:timeline}
    \vspace{-0.2in}
\end{figure}

This LLM-side sensitivity matters because the balance between the tool phase and the LLM phase is bursty rather than steady during real agent execution.
Each session alternates between tool execution and LLM generation, and these alternations are not aligned across sessions.
As Fig.~\ref{fig:timeline} illustrates, the resulting per-step decode batch size and GPU load fluctuate substantially, so the LLM engine repeatedly moves between under-utilization and overload.

Uncoordinated parallelization therefore does not always reduce task-level latency: correct speculation shortens exposed tool time, but blind speculation can make many sessions become ready for LLM generation earlier and closer together, increasing active batch size and creating phase imbalance.
We confirm this effect with a controlled experiment that accelerates tool execution by 2$\times$ while leaving the LLM-side scheduler unchanged; under high load, the extra returning workload slows LLM generation enough to absorb the tool-side gain, and in the worst case makes end-to-end latency worse than the baseline.
Therefore, the system must jointly decide when to execute speculative tool calls and how to pace sessions that return from tool-side into the LLM batch, avoiding phase imbalance that converts tool acceleration into LLM slowdown.

\subsection{Why Existing Systems Are Insufficient}

Existing LLM-serving and agentic-serving systems are tool-unaware and therefore do not address this joint problem.

Agent-level schedulers and general LLM serving engines optimize admission, batching, decoding, or session scheduling~\cite{kwon2023efficient, agrawal2024sarathiserve, duan2024muxserve, qiao2024conserve}.
However, they still treat ready LLM requests as the main scheduling unit and tool execution as an opaque blocking interval.
They do not know which sessions are waiting in tools, when tools will return, whether the tool side is congested, or the context lengths and KV footprints of returning sessions.
Therefore, they cannot smooth active LLM load or pace sessions before they re-enter the LLM engine.

KV cache swap and KV cache management systems optimize memory pressure~\cite{pan2025kvflow, kim2025kvzip, cai2025rkv, yang2025kvlink, prabhu2024vattention}, which is useful when GPU memory capacity limits admission.
However, our timeline experiment shows another common regime for agent serving: SM utilization and compute pressure are high, while GPU memory utilization is not full.
In this regime, KV cache management alone cannot substantially reduce LLM latency growth, because the bottleneck is not simply memory capacity: long-horizon agent tasks spend most of their latency outside the prefill stage.
Thus, LLM-side systems need tool-side state to control future LLM load.




\section{System Overview}
\label{sec:overview}

\sys is a coordinated tool--LLM parallelization system for agent serving.
It uses predictable future tool use to move concrete tool execution earlier in physical time, while pacing sessions that return to the load-sensitive LLM engine.
The key design principle is that reducing exposed tool time is useful only when the serving system preserves the gain instead of converting it into extra LLM queueing or decoding time.

\subsection{Design Goals}
\label{sec:overview-goals}

The Motivation leads to three design goals.

\vspace{0.35\baselineskip}
\noindent\textbf{G1: Predict concrete future invocations.}
\sys must use recurring control-flow patterns and implicit data-flow dependencies to predict executable future tool invocations, including arguments, rather than only tool names.

\vspace{0.35\baselineskip}
\noindent\textbf{G2: Non-interference speculative execution.}
Speculation must be interference-free along two dimensions.
First, on the resource side, speculative tool execution from many concurrent sessions must not delay authoritative tool invocations in any session.
Second, on the correctness side, speculative execution must not expose side effects or speculative results to the agent before the LLM confirms the invocation.
\sys therefore admits speculation selectively, bounds or preempts speculative jobs under contention, and keeps speculative results outside authoritative session state until matched.

\vspace{0.35\baselineskip}
\noindent\textbf{G3: Couple speculation with LLM-side pacing.}
Earlier tool completion can make sessions return to the LLM engine sooner and in tighter bursts, so \sys must pace returning LLM turns to keep active LLM load near the task-optimal region.

\subsection{System Architecture: Coordinated Tool--LLM Parallelization}
\label{sec:overview-architecture}

\begin{figure}[t]
    \centering
    \includegraphics[width=\linewidth]{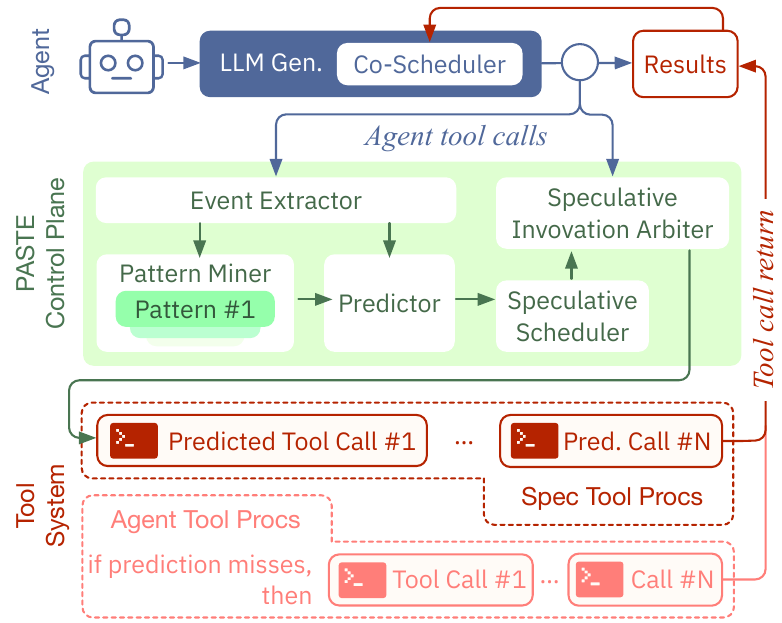}
    \caption{System architecture of \sys. \sys places a session-level control plane across the agent runtime, tool executor, and LLM serving engine.}
    \label{fig:system}
    \vspace{-0.12in}
\end{figure}

Figure~\ref{fig:system} shows the architecture of \sys.
Rather than treating the LLM engine and tool executor as independent services, \sys interposes a session-level control plane across the agent runtime, tool executor, and LLM serving engine.
Each agent request becomes an event stream with LLM turns, tool calls, tool outputs, timing signals, and session metadata.
The stream links the semantic LLM--tool loop to the physical serving timeline: \sys does not change which action the agent takes, but changes when predictable tool work runs and when completed sessions re-enter the LLM engine.

\noindent\textbf{Pattern Analyzer for concrete future invocations.}
The Pattern Analyzer converts recurring session structure into speculative invocation candidates.
It matches recent session history against recurring sub-workflows and, when the needed inputs are already present, instantiates the arguments required to execute the predicted invocation.

\noindent\textbf{Tool Speculation Scheduler for selective, interference-free overlap.}
The Tool Speculation Scheduler controls the speculative lifecycle.
It admits candidates according to confidence, expected benefit, and resource availability.
Authoritative tool invocations remain correctness-critical and keep priority.
Speculative work is bounded, deprioritized, or preempted under resource contention.
Speculative results stay outside authoritative session state until the LLM emits a matching invocation.
Tools with side effects require policy approval or a non-mutating variant.

\noindent\textbf{LLM--Tool Co-Scheduler for LLM-side pacing.}
The LLM--Tool co-scheduler paces sessions returned from the tool side to the LLM side.
It schedules LLM calls based on both authoritative and speculative tool execution signals, so that the active LLM load is optimal.

Together, these components create two complementary effects.
Intra-session overlap hides future tool latency, while cross-session pacing keeps the LLM engine in a latency-efficient region.

\subsection{Core Design Mechanisms}
\label{sec:overview-mechanisms}

This section summarizes the three core mechanisms; System Design details appear in \S\ref{sec:design}.

\subsubsection{Pattern-Aware Intra-Session Speculation}
\label{sec:overview-pattern-spec}

This mechanism creates overlap from predictable agent structure.
\sys learns recurring sub-workflows and implicit argument dependencies from prior sessions, then watches each live session for similar contexts.
When the required inputs already appear in the current session history, the Pattern Analyzer emits a concrete predicted invocation that can run while the LLM is still generating.
The design separates stable structure from volatile content: a search--visit pattern predicts that a web-fetch call is likely, but the concrete URL is copied from the current search output rather than guessed.

\subsubsection{Non-Interference Speculation Lifecycle}
\label{sec:overview-spec-lifecycle}

The speculation lifecycle separates best-effort predicted work from the authoritative path produced by the LLM.
A speculative invocation may start early, but it is not an agent-visible action until the LLM emits a matching authoritative invocation.
If the prediction matches, \sys reuses a completed result or promotes an in-flight job; if it misses, the work is discarded without changing session state.

\sys enforces two forms of non-interference.
\textbf{Resource side.} \sys prevents speculative jobs from delaying authoritative tool execution across concurrent sessions: authoritative jobs keep priority, and speculative jobs use only bounded or opportunistic capacity.
\textbf{Correctness side.} \sys prevents speculation from changing the agent outcome before confirmation: speculative results stay outside authoritative session state, and tools with side effects require an explicit safe-speculation policy or a non-mutating transformed execution.

\subsubsection{Joint LLM--Tool Co-Scheduling}
\label{sec:overview-coscheduling}

Joint LLM--tool co-scheduling converts local tool overlap into end-to-end latency reduction.
Speculation changes not only tool completion time, but also when sessions return to the LLM engine.
Without pacing, many sessions can become ready for generation at once, turning saved tool time into extra LLM queueing or decoding.
The LLM--Tool Co-Scheduler therefore observes tool completions, promoted jobs, active LLM requests, decode batch pressure, context length, and KV pressure.
It admits returning LLM turns so that exposed-tool-time reductions are preserved while the LLM engine stays near its task-optimal operating region.

\section{System Design}
\label{sec:design}

\sys realizes the overview design as a middleware/proxy around the agent's tool-dispatch path and LLM generation API, with scheduler hooks around the LLM serving engine.
This placement exposes the online event stream and per-session state used by three mechanisms: concrete prediction, interference-free speculative tool execution, and gain-preserving LLM-side pacing.

\subsection{Pattern Analyzer}
\label{sec:design-pattern-analyzer}

The Pattern Analyzer turns recurring agent-session structure into executable speculative invocation candidates.
The important point is that \sys does not only predict the next tool type; it predicts a concrete future invocation whose arguments can be derived from the current session history despite uncertainty in online agent execution.

\noindent\textbf{Event stream and representation.}
The event extractor records LLM turns, tool calls, tool outputs, timing signals, and session metadata from \sys's middleware/proxy event stream.
It normalizes each event into two parts: an event signature used for control-flow matching, and a payload retained for deriving concrete arguments.
Signatures intentionally capture stable control-flow information while excluding volatile natural-language content; payloads preserve the values that may later flow into predicted tool arguments.

\noindent\textbf{Prediction record.}
\sys stores each reusable prediction as a small pattern record.
The record contains a recent event context, a future tool target, an argument mapper, and an empirical confidence.
The context is matched over stable metadata such as tool type and success/failure status, while the mapper says where the future invocation's arguments should come from in the current session.

For example, after a successful search, a pattern may predict a web fetch and map its URL argument to the first URL in the current search result.
After a failed fetch, a fallback pattern can use the failure status and select the next URL from the same search result.
Thus the pattern captures recurring structure, while the live session supplies concrete values.

\noindent\textbf{Pattern pool construction.}
\sys builds the pattern pool in two passes.
First, it mines recurring short contexts over event signatures, which captures stable control flow while ignoring volatile natural-language payloads.
Second, it checks historical occurrences of each context to infer simple argument mappers, such as field lookup in structured tool output, choosing an indexed result with fallback, or lightweight string normalization.

Each candidate is validated against historical traces, and \sys keeps only patterns whose empirical probability is high enough for speculation.
Operator-supplied or offline-generated patterns can be added to the pool, but they pass through the same validation and confidence-estimation step before online use.

\noindent\textbf{Online prediction.}
For each live session, \sys maintains a bounded recent event window and matches the suffix of its signature stream against the validated pattern pool.
When a pattern matches, the analyzer applies the argument mapper to the current session's payloads.
This is late binding: the pattern decides what is likely to happen next, while the current session supplies the concrete values.
If the required payloads are already available, the analyzer emits a concrete invocation with tool name, canonicalized arguments, confidence, and expected latency benefit.
If only the tool identity or coarse parameters are known, the analyzer emits at most a preparation hint, not a fully executable tool call.
When multiple patterns match the same session state, the analyzer emits the corresponding hints without resolving conflicts locally; invocation-level arbitration is handled by the Tool Speculation Scheduler.
In both cases, prediction remains observational: the Pattern Analyzer forwards candidates to the Tool Speculation Scheduler and does not append anything to authoritative session state.

\subsection{Tool Speculation Scheduler}
\label{sec:design-tool-speculation-scheduler}

The Tool Speculation Scheduler moves concrete tool execution earlier in physical time while preserving the agent's semantic order.
Agent-issued tool calls enter the authoritative path exactly as before, while predicted calls enter a speculative path.
Both paths are launched through the same tool executor interface so that \sys can observe timing and reuse results, but speculative jobs and their outputs are isolated from authoritative session state until the LLM confirms the invocation.

\begin{figure}[t]
    \centering
    \includegraphics[width=0.88\linewidth]{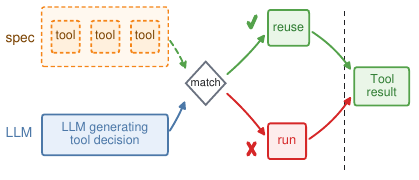}
    \vspace{-0.2in}
    \caption{Lossless speculative tool execution.}
    \Description{Workflow showing speculative tool calls running in parallel with LLM generation, followed by a match decision that either reuses the speculative result or runs the authoritative tool call normally before committing a tool result.}
    \label{fig:spectool}
    \vspace{-0.2in}
\end{figure}

Figure~\ref{fig:spectool} illustrates the execution rule.
While the LLM is still generating its next tool decision, the scheduler may start one or more predicted tool calls in the speculative path.
When the LLM later emits the authoritative invocation, \sys compares the canonicalized tool name and arguments against speculative jobs.
On a hit, the completed result is reused, or an in-flight job is promoted into the authoritative path; on a miss, the invocation simply falls back to normal execution.
The commit boundary in the figure is crossed only by an authoritative invocation, either through reuse or through normal execution.
Thus speculation is lossless: a wrong prediction can waste bounded work, but it cannot expose an unconfirmed result to the agent, change the session history, or change the final state.

\noindent\textbf{Admission.}
Speculation is admitted by policy plus runtime conditions, not by blindly launching every prediction.
The scheduler first deduplicates predictions at the invocation level, so multiple matching patterns do not launch redundant work.

\noindent\textbf{Admission checks.}
A candidate is launched only if it passes four checks.
It must be executable, meaning the tool name and arguments are known and can be canonicalized for later matching.
It must be policy-safe, either because the tool is side-effect-free or because the deployment provides a safe speculative variant.
It must have enough confidence and expected exposed-tool-time benefit to justify the work.
Finally, the speculative budget must have room, so speculative work does not consume resources reserved for authoritative invocations.

\noindent\textbf{Speculation level and priority.}
Fully parameterized, side-effect-free tools may run end to end.
Partial predictions are limited to preparation work such as warm-up or dependency loading.
Tools with side effects require an explicit safe-speculation policy or a non-mutating transformed execution, such as dry-run execution or execution against staging resources; otherwise they are not speculated.
Among eligible candidates, \sys greedily favors higher expected utility per unit resource, combining prediction confidence, tool time likely to be hidden, resource demand, and expected duration.

\noindent\textbf{Non-interference.}
Authoritative invocations keep the normal tool scheduling policy and priority.
Speculative jobs run only within bounded or opportunistic capacity, are lower priority, and can be suppressed or preempted when authoritative work needs resources.
When contention forces preemption, \sys reclaims speculative work before authoritative work and chooses lower-utility speculative jobs first.
This resource rule keeps speculative execution from delaying correctness-critical tool work across concurrent sessions, while the state-isolation rule keeps speculation from changing the agent-visible history before confirmation.

\noindent\textbf{Match and promotion.}
When an authoritative invocation arrives, \sys compares its tool name and arguments with completed and in-flight speculative jobs.
A completed match is reused directly.
An in-flight match is promoted to authoritative execution and becomes non-preemptible.
If no match exists, the authoritative invocation runs normally through the tool executor.

\noindent\textbf{Misses and signals.}
Every speculative job has one of four outcomes.
It is reused if it has completed and later matches an authoritative invocation, promoted if it is still running when the match arrives, discarded if no match arrives before expiry, or preempted if authoritative work needs the resources.
Only the first two outcomes can commit a result to authoritative session state.

The scheduler reports normal completions, reused results, promoted jobs, discarded jobs, preemptions, and the amount of exposed tool time saved to the LLM--Tool Co-Scheduler, which uses these cross-domain signals for LLM-side pacing.

\subsection{LLM--Tool Co-Scheduler}
\label{sec:design-llm-tool-coscheduler}

The LLM--Tool Co-Scheduler converts local tool overlap into task-level end-to-end latency reduction.
It is a separate scheduler from the Tool Speculation Scheduler: it consumes tool-side signals, but its decision point is LLM-side pacing.
Its central rule is to admit the ready LLM turn that preserves the largest expected exposed-tool-time reduction per unit of current LLM pressure.
This preserves tool-side gains while keeping the LLM engine in its task-optimal load region, instead of admitting every returned session as soon as its tool finishes.

\begin{figure}[t]
\centering
\includegraphics[width=\linewidth]{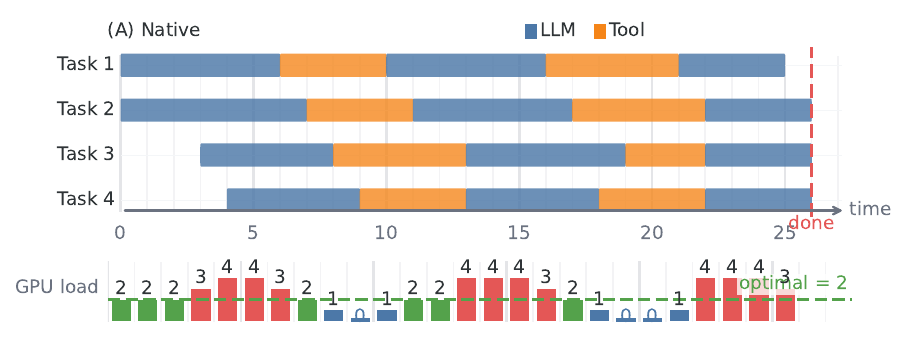}
\vspace{-0.05in}
\includegraphics[width=\linewidth]{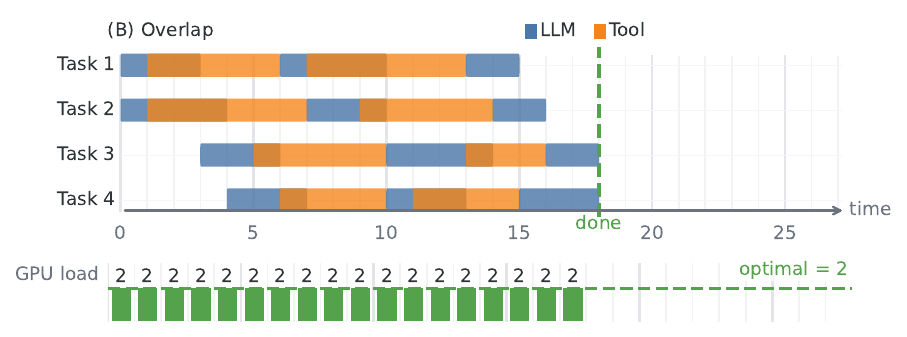}
\caption{Gain-preserving LLM--tool co-scheduling keeps LLM load near the task-optimal region.}
\Description{Two timeline panels compare native execution and overlapped execution for four tasks. Native execution finishes later and shows LLM GPU load fluctuating above and below the optimal load, while overlapped execution finishes earlier and keeps GPU load close to the optimal level.}
\vspace{-0.2in}
\label{fig:cosched-example}
\end{figure}

Figure~\ref{fig:cosched-example} makes the source of the gain concrete.
In the native schedule, each task alternates between LLM and tool phases independently, so aggregate LLM load repeatedly overshoots the optimal region and then drops into under-utilized gaps.
Shortening a tool wait alone would not necessarily reduce task latency, because the saved time can be converted into extra LLM queueing or slower decoding when many sessions return together.
In the co-scheduled timeline, tool waits are hidden behind useful LLM work from other sessions, and returning sessions are admitted only fast enough to keep the GPU near the task-optimal load.
The completion boundary therefore moves earlier not because an individual LLM or tool operation becomes faster, but because exposed tool time is removed without creating a compensating LLM-side bottleneck.

\noindent\textbf{Inputs and state.}
The co-scheduler consumes tool-side completion signals from authoritative execution, completed speculative jobs, and promoted in-flight speculative jobs.
These signals reveal how much exposed tool time has already been reduced and which sessions risk losing that gain to LLM queueing.
\sys also tracks each session's position in the LLM--tool chain: whether it is a new cold session, an active foreground session, a session returning from a tool, or a session likely to expose another tool wait after its next LLM turn.
For each ready turn, \sys estimates exposed-tool-time gain from recent service times, observed tool waits, pattern-derived next-tool likelihood, and session progress, while also tracking current LLM pressure.

\noindent\textbf{Pre-engine admission.}
The first control point is a pre-engine admission queue that decides which ready LLM turns are released into the serving engine.
The scheduler releases ready work to avoid GPU idle time when running concurrency is low, and softly gates new cold sessions once the engine has enough running work so the system does not spread capacity across too many partial sessions.
Among eligible turns, \sys ranks by gain efficiency:
\[
\mathrm{priority}(i) =
    \frac{\mathrm{ExposedToolGain}(i)}
         {\mathrm{LLMPressure}(i,\mathrm{load})}
    + \mathrm{Aging}(i).
\]
\texttt{ExposedToolGain} is the expected reduction in task-critical exposed tool time from admitting the turn now.
It has two concrete sources: realized gain from a completed or promoted tool result that can be consumed immediately, and future gain from reaching the next predictable tool wait early enough for speculation to hide it.
\texttt{LLMPressure} is the incremental pressure of admitting the turn now, including predicted LLM service time, current queue and batch pressure, context length, and KV/cache pressure.
\texttt{Aging} is a small fairness term.
Returned or promoted sessions are therefore not released merely because their tools finished; they are released when doing so preserves expected exposed-tool-time gain at acceptable LLM pressure.

\noindent\textbf{In-engine load shaping.}
The second control point is an in-engine load-shaping hook that keeps the running batch within the task-optimal load region after turns have been admitted.
The scheduler observes active requests, decode batch pressure, queue depth, context length distribution, and KV/cache pressure, and feeds these measurements back to update \texttt{LLMPressure}.
\sys summarizes the running batch with an engine-pressure score:
\[
\mathrm{EnginePressure}(B) =
    \mathrm{DecodeLoad}(B) + \gamma \cdot \mathrm{KVLoad}(B),
\]
\sys then keeps this score within a workload-aware pressure band:
\[
P_{\mathrm{low}} \le \mathrm{EnginePressure}(B) \le P_{\mathrm{high}}.
\]
Here $B$ is the current active or admitted batch; \texttt{DecodeLoad} captures active decoding work, while \texttt{KVLoad} summarizes context and KV-cache pressure.
This load shaping keeps the engine away from both under-utilization and overload, while preserving the serving engine's normal token-level scheduling inside the admitted set.
Together, pre-engine pacing and in-engine load shaping maximize preserved exposed-tool-time reduction under LLM load constraints, reducing partial-session work and avoiding the conversion of saved tool time into extra LLM queueing or decoding time.

\section{Implementation}
\label{sec:impl}

\sys is implemented as a non-invasive control plane around existing agent runtimes and a vLLM-based serving backend.
The prototype consists of 5k lines of TypeScript for agent-side integration and 2k lines of Python for vLLM-side integration.
On the agent side, \sys runs as a middleware/proxy on the tool-dispatch path and lightly wraps the LLM client for timing and session metadata.
Agents keep their original tool APIs; gemini-cli, Qwen Deep Research, and Virtual-Lab only require a small dispatch wrapper.

\noindent\textbf{Pattern Analyzer.}
The proxy records LLM turns, tool calls, outputs, timing signals, and session identifiers as an online event stream.
The analyzer normalizes events into compact signatures plus payloads, matches recent signatures against validated patterns, and binds matched patterns to current payloads.
Only fully instantiated predictions are forwarded; incomplete predictions remain preparation hints and never modify authoritative state.

\noindent\textbf{Tool Speculation Scheduler.}
The scheduler sends authoritative and speculative work through the same executor interface, so existing tools require no API changes.
Authoritative calls follow the normal path, while predictions enter a bounded speculative path.
\sys canonicalizes invocations so a later authoritative call can reuse a completed match, promote an in-flight match, or fall back to normal execution.
Misses expire outside authoritative state, while reuse and promotion events are reported to the LLM--Tool Co-Scheduler.

\noindent\textbf{LLM--Tool Co-Scheduler.}
The LLM-side component is a Python startup hook inside an unmodified vLLM server.
Task context is piggybacked on existing request metadata, including session progress, tool-wait information, and token estimates.
The hook updates per-session state and adjusts the order and admission of waiting LLM turns before native scheduling.
\sys does not change vLLM source code, serving APIs, or native execution paths; it only shapes which ready turns enter the engine.

\section{Evaluation}

This section evaluates \sys along seven axes: (1) End-to-End (E2E) latency improvement from jointly accelerating tool execution and LLM serving, (2) tool-side latency and throughput improvement, (3) where the improvement comes from (time breakdown), (4) scalability under concurrent sessions and bursty arrivals, (5) ablation of tool-side and LLM-side scheduling, (6) prediction quality, and (7) speculation overhead and safety under mispredictions.
Across all experiments, we hold hardware, LLM configuration, tool interfaces, and resource budgets constant to isolate system effects.

\subsection{Evaluation Methodology}
\label{sec:eval:method}

\paragraph{Experimental setup and controls.}
Unless stated otherwise, all systems run on the same hardware/software stack (Table~\ref{tab:eval:setup}).
We apply the same per-task timeouts and retry policy to all systems.
For pattern prediction, we mine tool-call patterns from a corpus of historical tasks and evaluate on a disjoint set of new tasks (no train/test overlap).

\paragraph{Workloads and baselines.}
We evaluate three agents: (i) \textbf{VirtualLab}~\cite{Swanson2025VirtualLab}, a science-focused agent designed to support end-to-end research workflows; (ii) \textbf{Qwen Deep Research}~\cite{alibabaNLP_deepresearch_repo}, an agent tailored for \emph{deep research}---multi-step, tool-assisted information gathering, synthesis, and report writing over open-ended questions; and (iii) \textbf{gemini-cli}~\cite{GeminiCLIBuild2025}, an open-source CLI-style agent for code modification as well as general-purpose tasks.
We use three benchmark families: \textbf{DeepResearchBench}~\cite{deep-research-bench} for deep-research-style tasks, \textbf{SWE-bench}~\cite{swebench} for software engineering and code-writing tasks, and \textbf{ScholarQA}~\cite{scholarQA} as a domain-specific scientific research benchmark.

Because \sys accelerates both sides of the LLM--tool loop, we compare against baselines that isolate the two sources of latency.
\textbf{vLLM}~\cite{kwon2023efficient} runs the unmodified agent on a high-throughput local LLM serving engine, capturing the performance of a strong but agent-unaware LLM backend.
\textbf{Agentix}~\cite{luo2025autellix} is an agent session-aware serving baseline that reduces task-level E2E latency through scheduling across agent sessions.
For tool-side acceleration, we compare against \textbf{ORION}~\cite{orion}, which reduces serverless tool startup latency through sizing, bundling, and prewarming, and \textbf{SpecFaaS}~\cite{stojkovic2023specfaas}, which speculatively executes serverless functions.
Both ORION and SpecFaaS use vLLM as the underlying LLM serving engine, so their differences from vLLM come from tool-side acceleration rather than a different model backend.

\paragraph{Real-world trace-driven request arrivals.}
We replay a production Azure Functions invocation trace~\cite{254430} to generate realistic, bursty arrivals.
Each trace record becomes one agent request issued at its logged timestamp, preserving the \emph{arrival process}.

\paragraph{LLM configuration.}
As Table~\ref{tab:eval:setup} shows, all LLM invocations are served locally with Qwen-DeepResearch-30B and Qwen3-30B-A3B. The first model is used for deep research tasks. The other is used for coding and scientific tasks.

\paragraph{Metrics.}
Our primary metrics are \textbf{E2E latency} (request arrival to final agent response) and observed tool execution latency.
We also report tail latency (p95/p99), \textbf{tool stall time} (time the agent waits for tool execution results), LLM-side queueing time, throughput, speculative \textbf{hit rate}, and resource overhead.

\begin{table}[t]
  \centering
  \caption{Evaluation setup}
  \label{tab:eval:setup}
  \vspace{-0.1in}
  \small
  \begin{tabular}{p{0.32\columnwidth}p{0.58\columnwidth}}
    \toprule
    \textbf{Component} & \textbf{Setting} \\
    \midrule
    Cluster & 4 nodes \\
    Per-node CPU & 96 vCPUs, AMD EPYC 7V13  \\
    Per-node Memory & 512 GB \\
    Per-node GPU & 8 GPU, NVIDIA A100 80GB  \\
    Total GPUs & 32$\times$ NVIDIA A100 \\
    \midrule
    LLM Serving Engine & vLLM \\
    Models & Qwen-DeepResearch-30B~~\cite{tongyidr}, Qwen3-30B-A3B~~\cite{qwen3_30b_a3b} \\
    \midrule
    \vspace{-0.1in}
  \end{tabular}
\end{table}

\subsection{E2E Latency Reduction}
\label{sec:eval:e2e}

\begin{figure}[t]
  \centering
  \includegraphics[width=\columnwidth]{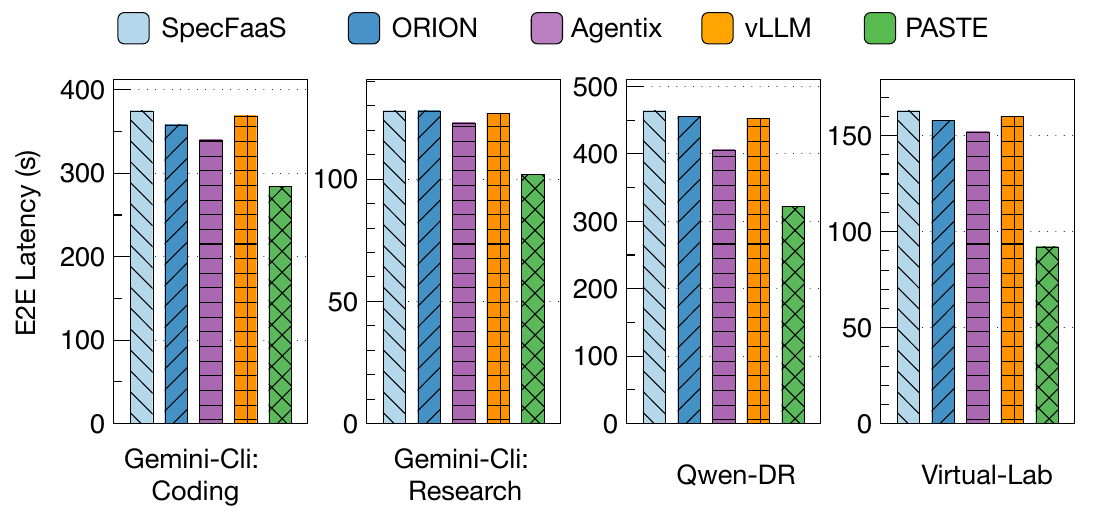}
  \vspace{-0.3in}
  \caption{Average E2E latency comparison against all baselines.}
  \vspace{-0.2in}
  \label{fig:eval:e2e}
\end{figure}

We first quantify whether \sys reduces E2E latency for each agent.
Figure~\ref{fig:eval:e2e} summarizes the results.
Across all agents, \sys consistently reduces E2E latency relative to vLLM, Agentix, ORION, and SpecFaaS.
\sys reduces average latency by up to 43.5\%, with p99 tail latency improving by up to 55.4\%.
The improvement appears against both LLM-side baselines and tool-side baselines, indicating that neither serving-side scheduling nor tool-side acceleration alone removes the dominant stalls in the agent loop.

The improvement holds across both average and tail latency: the combination of speculative tool execution and LLM--tool co-scheduling reduces exposed tool stalls and lowers LLM-side waiting.
We further quantify this mechanism with a time breakdown in Sec~\ref{sec:eval:breakdown}.

\subsection{Tool-Side Acceleration}
\label{sec:eval:tool}

We next isolate \sys's ability to accelerate the tool side of agent execution.
Since vLLM and Agentix change the LLM-side serving policy but do not accelerate the tool execution path, this experiment compares \sys only with the two tool-side baselines, ORION and SpecFaaS.
Both baselines are backed by vLLM for LLM serving.

\begin{figure}[t]
  \centering
  \includegraphics[width=0.9\columnwidth]{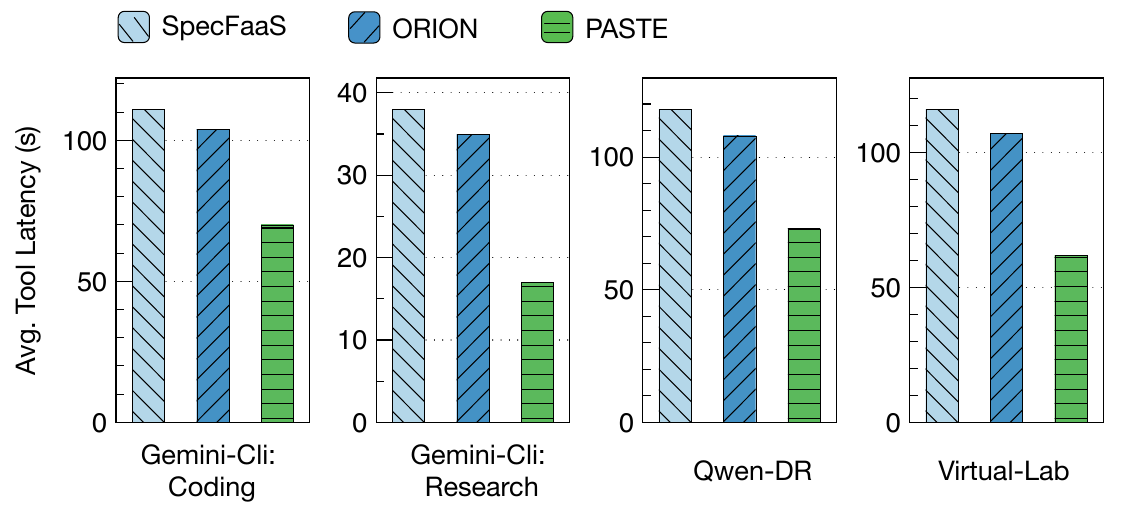}
  \vspace{-0.2in}
  \caption{Average tool latency comparison against tool-side baselines.}
  \label{fig:eval:tool}
  \vspace{-0.1in}
\end{figure}

\begin{figure}[t]
  \centering
  \includegraphics[width=\columnwidth]{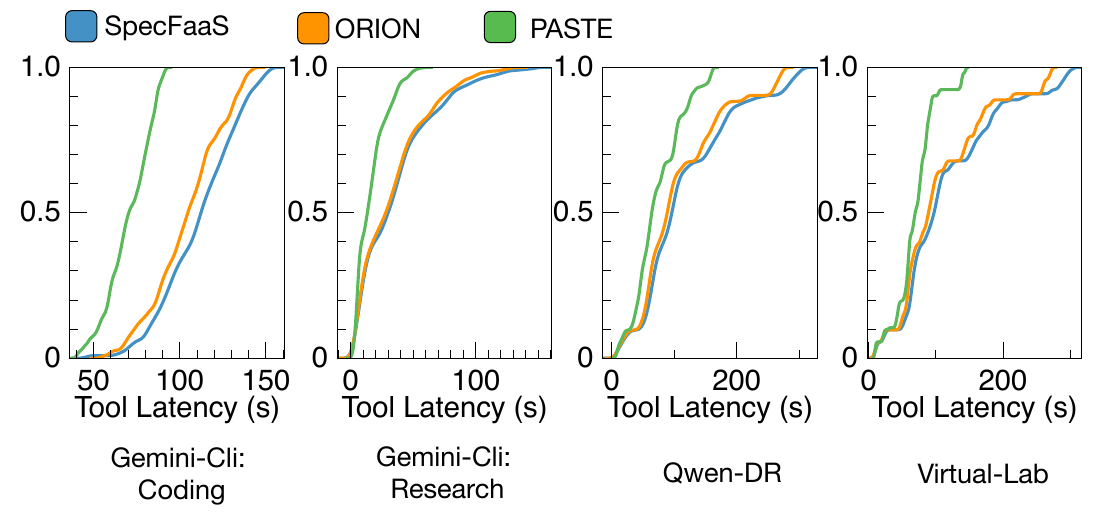}
  \vspace{-0.2in}
  \caption{CDF of tool latency for each task}
  \label{fig:eval:tool_cdf}

\end{figure}

Figure~\ref{fig:eval:tool} reports average tool latency aggregated across all agents, and Figure~\ref{fig:eval:tool_cdf} plots the corresponding per-task CDF.
Across agents and benchmarks, \sys significantly reduces observed tool latency relative to both ORION and SpecFaaS.
\sys reduces average tool latency by up to 55.2\% and p99 tool latency by up to 60.6\%.
Aggregated across configurations, the average tool-side speedup is 1.71$\times$ over ORION and 1.83$\times$ over SpecFaaS.
These gains are consistent with speculative execution making predicted tool results available before they appear on the exposed critical path.

\begin{figure}[t]
  \centering
  \includegraphics[width=\columnwidth]{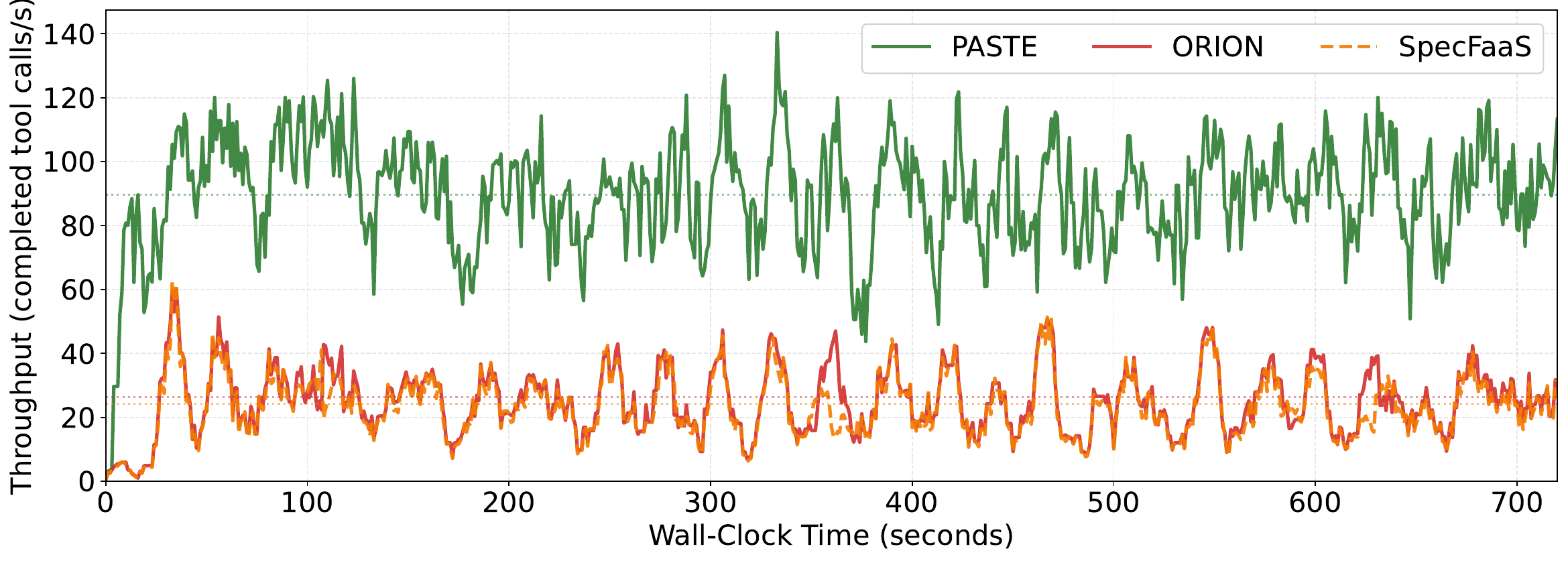}
  \vspace{-0.2in}
  \caption{Tool throughput under trace-driven arrivals.}
  \label{fig:eval:throughput}
  \vspace{-0.1in}
\end{figure}

Figure~\ref{fig:eval:throughput} shows that the latency reduction translates into higher completed-tool throughput under bursty arrivals.
\sys completes more tool work under the same trace-driven arrivals, while keeping authoritative tool calls on the high-priority path.

\begin{figure}[t]
    \centering
    \includegraphics[width=0.45\columnwidth]{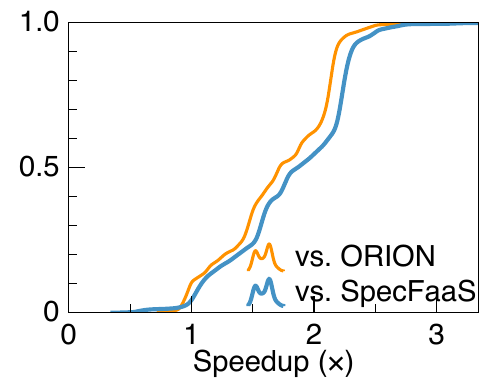}
    \vspace{-0.15in}
    \caption{CDF of per-request tool speedup over ORION/SpecFaaS.}
    \label{fig:speedup-cdf}
    \vspace{-0.2in}
\end{figure}

Finally, we aggregate all tasks and examine \sys's tool-side speedup relative to the tool baselines.
Figure~\ref{fig:speedup-cdf} reports the CDF of per-request tool speedup over ORION and SpecFaaS after pooling all requests.
Most requests observe positive speedup (97\% above 1$\times$), while the worst cases remain close to parity, consistent with \sys's low speculation overhead when predictions do not hit.

These gains are robust across agents, benchmarks, and LLM settings.
Improvements are largest for tool-heavy tasks where tool stalls dominate the critical path, but remain non-negative for more LLM-dominated tasks, indicating that \sys introduces low overhead when speculation does not hit.
Overall, speculative tool execution reliably converts a portion of tool-wait time into ready speculative results, improving both median and tail tool latency.

\subsection{Time Breakdown Analysis}
\label{sec:eval:breakdown}

To explain \emph{why} \sys reduces E2E latency, we instrument the runtime and attribute time to two components: the exposed tool-side path and the LLM-side serving path.
The tool-side path includes tool startup and execution time that remains visible on the task critical path, while the LLM-side path includes model execution and queueing before the next agent turn can run.

\begin{figure}[t]
  \centering
  \includegraphics[width=\columnwidth]{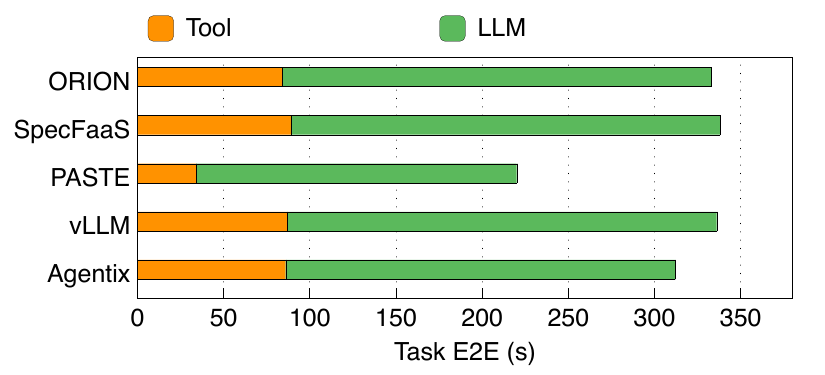}
  \vspace{-0.3in}
  \caption{Time breakdown across all baselines.}
  \label{fig:eval:breakdown}
  \vspace{-0.2in}
\end{figure}

Figure~\ref{fig:eval:breakdown} reports the breakdown for vLLM, Agentix, ORION, SpecFaaS, and \sys.
\sys reduces both components rather than shifting cost between them: exposed tool-side time falls to 34~s from 84--89~s under the tool-side baselines, while the LLM-side component falls to 186~s from 226--249~s under the LLM-side baselines.
This corresponds to roughly 60\% less exposed tool time and 18\%--25\% less LLM-side time.

These results confirm that \sys's speedups come from both sides of the loop.
ORION and SpecFaaS improve the tool path but leave LLM serving unchanged, while Agentix reduces LLM-side waiting but does not hide exposed tool stalls.
By combining speculative tool execution with LLM--tool co-scheduling, \sys reduces the two visible sources of latency at the same time.

\subsection{System Scalability}
\label{sec:eval:scalability}

We next evaluate scalability to determine whether \sys maintains low latency under many concurrent agent sessions, and to verify that speculation does not harm isolation by increasing queueing for authoritative tool calls.
We stress the system by sweeping arrival rate and concurrent sessions while holding resource budgets fixed.
We compare scalability against vLLM and Agentix because the multi-session bottleneck is the LLM serving engine: tool-side acceleration has limited impact as concurrency increases, while LLM queueing dominates task-level latency.

\begin{figure}[t]
  \centering
  \includegraphics[width=\columnwidth]{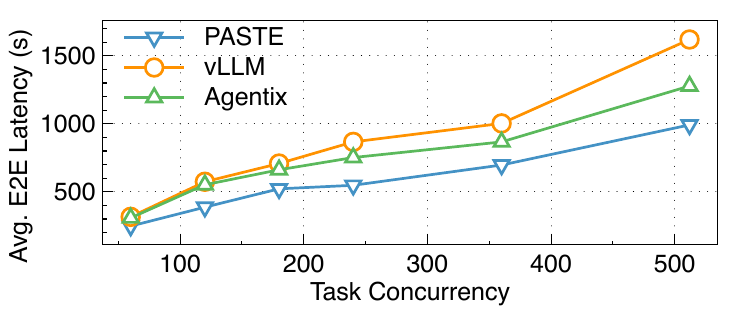}
  \vspace{-0.3in}
  \caption{Scalability under multi-session concurrent agent requests: speedup compared with LLM-side baselines.}
  \label{fig:eval:scalability}
  \vspace{-0.1in}
\end{figure}

Figure~\ref{fig:eval:scalability} summarizes \sys's E2E speedup performance as workload increases.
At each concurrency, \sys sustains at least 1.27$\times$ speedup over vLLM and 1.24$\times$ over Agentix.
Pooled across the sweep, the corresponding speedups are 1.50$\times$ and 1.30$\times$.
These results demonstrate that \sys scales without violating isolation: speculative work is throttled by explicit budgets and remains preemptible, so it does not crowd out authoritative tool execution.

\subsection{Ablation Study}
\label{sec:eval:ablation}

We isolate the contribution of each scheduling component by comparing \sys with two ablated variants and two agent/LLM serving baselines.
\textbf{PASTE-Tool-Only} enables tool-side acceleration through speculative tool execution, but disables the LLM--Tool Co-Scheduler; tool-returning sessions therefore re-enter the LLM side without gain-aware pacing.
\textbf{PASTE-LLM-Only} disables speculative tool execution and only uses first-come-first-served LLM admission to reduce excessive GPU load.
We also compare against vLLM~\cite{kwon2023efficient} and Agentix~\cite{luo2025autellix}.

\begin{figure}[t]
  \centering
  \includegraphics[width=0.9\columnwidth]{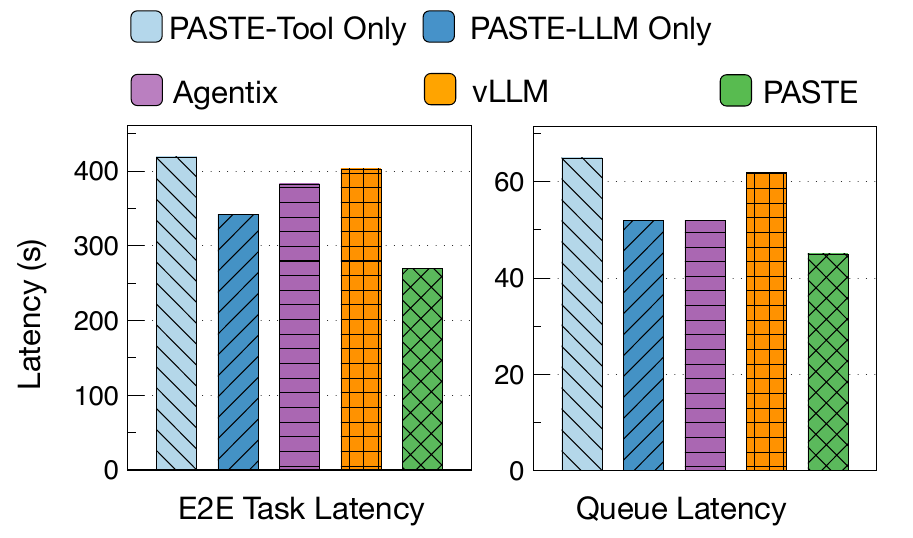}
  \vspace{-0.1in}
  \caption{Task's E2E latency in ablation evaluation.}
  \label{fig:eval:ablation}
  \vspace{-0.1in}
\end{figure}

Figure~\ref{fig:eval:ablation} shows that both sides are necessary.
PASTE-Tool-Only performs worst: although tool speculation shortens tool execution, it also makes sessions return to the LLM more frequently and in tighter bursts, increasing average LLM queueing time to 65~s and raising E2E latency to 419~s.
PASTE-LLM-Only reduces queueing pressure, but without speculative tools it cannot hide exposed tool time, so E2E latency remains 342~s.
Full \sys combines the two mechanisms: speculative tools reduce exposed tool stalls, while LLM--tool co-scheduling preserves this gain before it turns into GPU queueing.
As a result, \sys reduces E2E latency to 270~s and LLM queueing time to 45~s, cutting E2E latency by 21.1\%--35.6\% and queueing time by 13.5\%--30.8\% across the ablations and serving baselines.

\subsection{Pattern Prediction Accuracy}
\label{sec:eval:accuracy}

Speculative execution is only effective when the system can correctly anticipate near-future tool calls.
We therefore measure prediction quality using \textbf{Top-1 accuracy} (the highest-probability predicted tool matches the next tool actually invoked) and \textbf{Top-3 recall} (the next tool appears among the three most likely predictions).
Top-3 recall is particularly relevant because \sys may speculatively execute multiple candidates within a bounded budget.
We also report \textbf{overall hit rate}, defined per tool call as the probability that \emph{any} speculatively executed prediction matches the next tool actually invoked.
Because \sys can issue multiple predictions (and increase the number of candidates when the system is lightly loaded and spare resources are available), overall hit rate can be high even when Top-1 accuracy is low.

\begin{figure}[t]
  \centering
  \includegraphics[width=0.85\columnwidth]{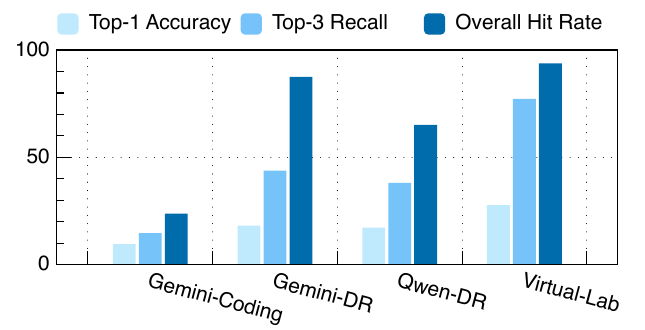}
  \vspace{-0.1in}
  \caption{Pattern prediction quality: Top-1 accuracy, Top-3 recall, and overall hit rate.}
  \vspace{-0.1in}
  \label{fig:eval:accuracy}
\end{figure}

Figure~\ref{fig:eval:accuracy} reports prediction quality by benchmark family.
Overall, the predictor achieves up to 27.8\% Top-1 accuracy, 43.9\% Top-3 recall, and 93.8\% overall hit rate.
Accuracy is higher for structured tool sequences (e.g., compilation/test loops) and lower for open-ended exploration patterns typical of broad research tasks.

Despite imperfect Top-1 accuracy, strong Top-3 recall is sufficient in practice: \sys can speculate on a small set of likely tools and still reduce exposed tool time when any candidate hits.
When prediction is uncertain, the explicit speculation budget bounds wasted work and prevents negative interference with authoritative execution.

\subsection{Side-effect Evaluation}
\label{sec:eval:sideeffects}

Finally, we evaluate safety to ensure that enabling speculation does not change externally visible behavior or violate isolation, even under mispredictions.
We audit speculative executions to measure (i) how many speculative actions would have caused external side effects, (ii) whether such effects were prevented from committing, and (iii) any divergences in final outputs relative to authoritative-only execution.
Across all workloads, \sys detects 602 potentially side-effecting speculative actions among over 20,000 speculative actions and prevents them from committing. No task produces a different final result compared with the baselines.
Overall, the results support \sys's safety claim: speculative actions are contained by policy and sandboxing and only become externally visible after authoritative confirmation.

\subsection{Resource Overhead}
\label{sec:eval:efficiency}

Resource overhead remains small relative to the latency savings.
At a moderate speculation budget, \sys reduces tool execution latency by 48\% using 1--3 otherwise idle CPU cores and 250 MB of additional memory.
Normalized by latency saved, this costs 0.02 core-seconds of CPU, 2.6 MB of memory, and 0.9 MB of network bandwidth per second of latency reduction.
These results show that \sys can be tuned to a practical ``sweet spot'' in which most speculative work is converted into useful latency reduction, while the remaining waste is bounded by explicit budgets.
Overall, PASTE is lightweight enough to be deployed as a sidecar alongside existing agent runtimes, delivering latency wins without requiring dedicated infrastructure or disruptive changes to the execution stack.
The pattern predictor and scheduling policy add less than 100 ms of latency overhead.

\section{Related Work}

\textbf{LLM and agentic serving.}
General LLM serving systems improve throughput and latency through scheduling, batching, multiplexing, distributed serving, and KV/context memory management~\cite{kwon2023efficient,agrawal2024sarathiserve,wu2023fast,duan2024muxserve,qiao2024conserve,qiu2024efficient,li2025throughput,pan2025kvflow,kim2025kvzip,cai2025rkv,yang2025kvlink,hu2024epic,xie2025strata,prabhu2024vattention}.
Agentic serving systems extend this line to tool-using, multi-turn programs with workflow/session-aware execution, intercept support, and cache management~\cite{luo2025autellix,parrot24,zhang2026megaflow,giusti2025federation,zhang2025ufo3,guo2025agenticlybic,zheng2024sglang,abhyankar2024infercept,wang2025augserve,zhai2026toolcaching,xu2025alignment,li2026continuum,chen2026concur,kang2026thunderagent,zheng2026efficient,xia2026idleness}.
However, this serving line remains largely tool-execution unaware: it does not directly speed up tool calls, and it misses opportunities to use tool-side information to further optimize LLM serving, which \sys{} targets.

\textbf{Tool \& runtime acceleration.}
Tool execution performance is shaped by serverless/workflow runtimes, so recent work reduces \emph{workflow orchestration overhead} and mitigates \emph{cold starts} via better provisioning/worker reuse~\cite{liu2023unum,joosen2025serverless,liu2024jiagu,prewarm}.
Additional systems target \emph{cross-service caching/state management} for microservice graphs and serverless settings~\cite{zhang2024mucache,zhang2025causalmesh}, and speed up tool-heavy pipelines with more efficient \emph{data passing/transfer paths}~\cite{marcelino2024truffle,wu2024faastube,marcelino2025databelt}.
Yet, most of these techniques assume a largely static workflow/microservice graph and cannot directly exploit the online, LLM-generated control flow and context-dependent argument binding in agent loops, which \sys{} is designed to handle.

\textbf{Speculation.}
Speculation has been applied to \emph{serverless functions} to accelerate workflows by executing likely-needed tasks early~\cite{stojkovic2023specfaas}.
Emerging agentic work explores \emph{speculative actions and speculative tool calls} to overlap tool latency with reasoning/search while preserving correctness constraints~\cite{ye2025speculative,huang2025reducing,nichols2025spectoocalls}. Compared to \sys{}, existing speculation mechanisms struggle to (i) infer the concrete, context-dependent arguments that are only produced online by the agent, making accurate tool-call speculation difficult, and (ii) manage the correctness and side-effect risks of executing wrong calls, which requires explicit, risk-aware control.

\section{Conclusion}

In this paper, we presented \sys, a coordinated tool--LLM parallelization system for low-latency LLM agent serving.
\sys predicts concrete future tool invocations from recurring agent patterns, executes them speculatively while the LLM generates, and paces returning sessions so that hidden tool time is not converted into extra LLM-side delay.
Across deep research, software engineering, and scientific-agent workloads, \sys reduces exposed tool stalls and LLM-side waiting while preserving the agent's semantically sequential execution.

\bibliographystyle{ACM-Reference-Format}
\sloppy
\bibliography{references}

\end{document}